\documentclass[a4paper]{jpconf}
\usepackage{graphicx}
\begin{document}
\title{Fundamental Physics from the Sky: Cosmic Rays, Gamma Rays and the Hunt for Dark Matter\footnote{Invited plenary talk, 18th International Symposium on Particles, Strings and Cosmology (PASCOS), Merida, Mexico, June 3-8, 2012.}}

\author{Stefano Profumo}

\address{Santa Cruz Institute for Particle Physics, Santa Cruz, CA 95064, USA, and\\ Department of Physics, University of California,
1156 High St., Santa Cruz, CA 95064, USA}

\ead{profumo@ucsc.edu}

\begin{abstract}
Can we learn about New Physics with astronomical and astro-particle data? Understanding how this is possible is key to unraveling one of the most pressing mysteries at the interface of cosmology and particle physics: the fundamental nature of dark matter. I will discuss some of the recent puzzling findings in cosmic-ray electron-positron data and in gamma-ray observations that might be related to dark matter. I will argue that recent cosmic-ray data, most notably from the Pamela and Fermi satellites, indicate that previously unaccounted-for powerful sources in the Galaxy inject high-energy electrons and positrons. Interestingly, this new source class might be related to new fundamental particle physics, and specifically to pair-annihilation or decay of galactic dark matter. This exciting scenario is directly constrained by Fermi gamma-ray observations, which also inform us on astrophysical source counterparts that could also be responsible for the high-energy electron-positron excess.  Observations of gamma-ray emission from the central regions of the Galaxy as well as claims on a gamma-ray line at around 130 GeV also recently triggered a wide-spread interest: I will address the question of whether we are really observing signals from dark matter annihilation, how to test this hypothesis, and which astrophysical mechanisms constitute the relevant background.
\end{abstract}

\section{Introduction}
The standard picture for galaxies such as our own Milky Way is one where the visible matter is embedded in a halo of {\em dark matter} of as of yet unknown particle nature. Many theoretically well-motivated particle theories beyond the Standard Model predict the existence of particles that could be the dark matter. In many of these theories, such particles can pair-annihilate, producing, as a result, Standard Model particles that can be detected. These detectable final states include antimatter particles, high-energy neutrinos, gamma rays, electrons and positrons ($e^\pm$), and secondary products of $e^\pm$ energy losses, such as radiation at X-ray frequencies from Inverse Compton and at radio frequencies from synchrotron.

Collectively, the search for a signature from dark matter looking at its annihilation products is known as {\em indirect detection}. The key question for a particle physicist is: can we do fundamental physics with indirect dark matter detection? Specifically, I will address the sub-question of whether this is possible employing cosmic ray $e^\pm$ and gamma rays.

It is relevant to note that, historically, big leaps in our understanding of the underlying structure of particle physics models have been made possible by cosmic ray studies. These leaps include the discovery of antimatter, of a second generation of matter fermions, of mesons, and, more recently, of neutrino masses and mixing.

Indirect dark matter detection is a multi-wavelength and multi-messenger endeavor. Key to doing fundamental physics is to control astrophysical backgrounds to signatures possibly stemming from dark matter. Not surprisingly, the most exciting places to look for a dark matter signature correspond to some of the places with the ``worse'' astrophysical background. In this contribution, I address the high-energy positron excess, dark matter annihilation in the Galactic Center, and the 130 GeV line.

\section{The high-energy positron excess}
The Pamela telescope has observed a significant excess of positrons in the 10-100 GeV range, well above the standard Galactic cosmic-ray expectations \cite{pamela}. A notable recent result is the confirmation, by the Fermi Large Area Telescope (LAT), of this excess \cite{fermiepem}, a result which extends the rise in the positron fraction (the fraction of positrons to electrons plus positrons as a function of energy) up to $\sim$200 GeV. The notable experimental technique employed by the Fermi-LAT team hinges upon the use of the Earth shadow, as a function of the cosmic ray $e^\pm$ arrival direction and energy: For particular arrival directions on the detector, electrons or positrons of given energy are completely forbidden, their trajectories, bent by the geomagnetic field, being shadowed by the Earth. The result of Ref.~\cite{fermiepem} is notable from at least three standpoints:
\begin{enumerate}
\item It confirms the Pamela result, that was potentially plagued by positron mis-identification (for every 50 GeV cosmic-ray positron, one has about 10 electrons and $10^4$ protons, the latter being in principle easily mis-identified with positrons);
\item It extends the Pamela result to higher energy, with no evidence for a turnover or suppression in the source of high-energy positrons;
\item It rules out certain models for the positron fraction excess, for example supernova remnant inhomogeneities in the vicinity of the Sun \cite{Shaviv:2009bu}, predicting a turnover of the positron fraction between 100 and 200 GeV.
\end{enumerate}

The question of the nature of the additional source of big-energy $e^\pm$ remains to be addressed: numerous studies have advocated dark matter annihilation, despite serious model building issues, including:
\begin{enumerate}
\item The need for very large annihilation cross sections, $10^2-10^3$ times larger than the natural scale needed for a standard thermal freeze-out scenario to produce dark matter with the correct relic density;
\item The need for rather large dark matter masses (even larger after the Fermi $e^\pm$ result);
\item The need to suppress excess antiprotons and therefore to select very special annihilation final states, such as $\mu^+\mu^-$.
\end{enumerate}
Despite these shortcomings, several possible scenarios have been proposed. Any such scenario is constrained by the necessary occurrence of final state radiation and of inverse Compton from the $e^\pm$ needed in the annihilation final state to explain the positron fraction. One of the best handles to constrain dark matter explanations of the high-energy positron excess is observing nearby galaxy clusters at gamma-ray frequencies, including the case of dark matter decay \cite{clusterlim}. The size of clusters of galaxies is such that the energy loss time scale is much shorter than the time-scale associated to the diffusion of $e^\pm$ out of the system: clusters are thus perfect ``calorimeters'' for cosmic-ray $e^\pm$ in the 10-100 GeV range \cite{Colafrancesco:2005ji, Profumo:2008fy, Jeltema:2008vu}. It should be noted that the resulting limits depend on assumptions on substructure, but provide rather stringent constraints on the range of masses and cross sections that could explain the $e^\pm$ excess \cite{clusterlim}. Additional constraints come from observations of the cosmic microwave background and of the diffuse isotropic extragalactic gamma-ray background (see e.g. \cite{Profumo:2009uf}).

An alternative scenario, proposed among other studies in Ref.~\cite{occam, grassoetal}, postulates that the excess $e^\pm$ originate from existing nearby radio pulsars, in certain ranges of age and distance, and for plausible $e^\pm$ injection efficiencies. The Fermi-LAT discovery of multiple radio-quiet pulsars has further strengthen the pulsar scenario, providing a number of previously unknown nearby pulsars that likely contribute to the local cosmic-ray $e^\pm$ population \cite{gendelev}. The pulsar interpretation of the $e^\pm$ thus remains a viable and probable explanation to the cosmic-ray positron excess, which is, in particular, compatible with the new Fermi determination of the positron fraction \cite{fermiepem}.

Can we discriminate between dark matter and pulsars as the origin of the $e^\pm$ excess? A smoking gun for the pulsar interpretation is the detection of an anisotropy in the arrival direction of the cosmic ray $e^\pm$, especially if such anisotropy pointed towards a known pulsar \cite{fermianiso}. For dark matter, the general expectation is a high-latitude inverse Compton emission from annihilation of dark matter in the halo, with a peculiar spectrum. None of these smoking guns has been observed so far, although limits on the anisotropy of the $e^\pm$ arrival directions are close to the level expected if the excess were due to one single nearby pulsar \cite{fermianiso}. At present, both interpretations are compatible with data, and the origin of the positron excess remains still unsettled (albeit with increasingly tight constraints on the dark matter explanation).

Looking at the nearby future, the Alpha Magnetic Spectrometer AMS-02 \cite{ams}, successfully launched and deployed on the international Space Station, is undergoing calibration and alignment. The expectation from AMS-02 is to deliver preliminary data on the positron fraction in early 2013, with the next priorities antiproton data and high-Z antimatter searches \cite{iris}.

\section{The Galactic Center}
Multiple studies have identified a possible signal from dark matter annihilation at the center of the Galaxy using Fermi-LAT data (see e.g. \cite{hoopergood, goodhooper, lindenhooper}). The results of these studies illustrate how crucial a proper modeling of the diffuse emission and of the emission from the very center of the Galaxy is to reliably pinpoint an ``anomalous'' signal from dark matter. An {\em under-fitting} of the astrophysical background leads to the possibly misleading identification of excesses to be attributed to dark matter. The opposite issue is the {\em over-fitting} of the diffuse gamma-ray model, that would lead to fine-tuning the astrophysical background to incorporate at all costs any possible excess.

A key element in searching for dark matter in the Galactic center is understanding the gamma-ray source associated with Sgr A*. This source, if fueled by protons, can be responsible for a significant diffuse emission from the protons accelerated by the central supermassive black hole in the innermost Galactic center region, which is of crucial importance to look for a signal from dark matter annihilation. A careful treatment of the target density of gas in the region shows that a key diagnostic to unveil whether Sgr A* accelerates protons is the gamma-ray source morphology \cite{lindenlovegroveprofumo}. Only future Cherenkov Telescope Arrays will have the needed angular resolution to pinpoint a hadronic origin for Sgr A*, leading to possibly understanding some of the diffuse gamma-ray emission in the region \cite{lindenprofumo}.

If the dark matter interpretation of the diffuse excess in the Galactic center is correct, a likely mass range is about 10 GeV. This mass range resonates with a series of recent direct and indirect dark matter detection results (notably, recently, at radio frequencies \cite{radiofil, arcade2}). Albeit there is no conclusive evidence for new physics for any of these signals, this is an interesting possibility \cite{hooper10gev} worth investigating on a case-by-case basis and definitely worth model-building efforts (for recent examples see e.g. \cite{profumoboucenna, hooper10gev}).

\section{the 130 GeV line}
Evidence for a monochromatic gamma-ray line at 130 GeV has emerged from the analysis, in Ref.~\cite{weniger}, of optimized signal-to-noise regions in the gamma-ray sky, at about the 3-$\sigma$ level, once the look-elsewhere effect is taken into account \cite{weniger}. In Ref.~\cite{lineprofumolinden}, myself and Linden pointed out that the relevant region of interest overlaps with the so-called Fermi bubbles \cite{fermibubbles}, which are thus an important source of background to the line. It was also noticed that a broken power-law spectrum (as what observed in the Fermi bubbles \cite{fermibubbles}) could be mis-identified, or could create confusion, with a gamma-ray line \cite{fermibubbles}. 

It is important to note that the outline of the relevant regions for the Fermi bubbles only partly overlaps with the region where most of the 130 GeV photons originate from. However, the Fermi bubble regions were specifically outlined outside the Galactic plane (to suppress the bright Galactic diffuse background), while the observed X-ray counterpart to the Fermi bubbles \cite{fermibubbles} indicates precisely that most of the emission associated with the mechanism fueling the bubbles should originate squarely from the Galactic center region where the 130 GeV line is most prominent.

It is currently unclear whether the 130 GeV line can be associated with an instrumental effect (for a recent study see e.g. \cite{Finkbeiner:2012ez}) --- perhaps issues with the energy-reconstruction algorithm employed by the current version of the Fermi Tools. Likely, the forthcoming ``Pass 8'' Fermi Tools software suite will address relevant aspects associated with instrumental systematics. Future observations with ground based Cherenkov Telescopes with a low energy threshold (e.g. H.E.S.S.-II and, further in the future, CTA) could also be decisive in deciding of the validity and nature of the 130 GeV line observation.

\section{Conclusions}
The search for dark matter is in an extraordinarily exciting phase:
\begin{itemize}
\item Direct dark matter experiments are tapping into the exciting low-mass region, and carving into interesting regions of particle dark matter parameter space;
\item The Large Hadron Collider at CERN is producing results that are sharpening the predictions of certain dark matter models (e.g. Universal Extra Dimensions) and could soon provide clues on the correct beyond-the Standard Model framework;
\item Results from AMS-02 will help understanding how to properly treat cosmic ray propagation, and will provide more information on cosmic ray antimatter, especially positrons, antiprotons and antideuterons;
\item Post-Fermi-LAT high-energy telescopes will cover the high-mass end of the particle dark matter parameter space (CTA), with the low-mass end possibly under scrutiny with the new hard X-ray facilities \cite{hardxrays}.
\end{itemize}

The indirect search for dark matter resonates with a famous quote by painter Ren\'e Magritte\footnote{I thank Pasquale Serpico for bringing this quote to my attention.}: ``Everything we see hides another thin, we always want to see what is hidden by what we see''. 

\section*{Acknowledgments}
\noindent  SP is partly supported by an Outstanding Junior Investigator Award from the US Department of Energy and by Contract DE-FG02-04ER41268, and by NSF Grant PHY-0757911.

\end{document}